\documentclass[aps,reprint,prl,superscriptaddress,showpacs,floats,floatfix]{revtex4-2}
\usepackage{morefloats}
\usepackage{mathptmx}
\usepackage{indentfirst}
\usepackage{amsmath,graphicx,dcolumn}
\usepackage[colorlinks,citecolor=green]{hyperref}
\usepackage[usenames]{color}
\usepackage{datetime}
\usepackage{textcomp}
\usepackage{ulem}
\usepackage{float}

\newcommand{\blue}{\textcolor{blue}}

\newcommand*{\mjcm}{$\mu$J/cm$^2$}

\begin{document}

\title{Ultrafast Phonon Hardening and Nonthermal Lattice Potential Reconstruction in the Charge-Density-Wave Material 1$T$-TiSe$_{2}$}

\author{Xue-Qing Ye}
\affiliation{School of Physics, Central South University, Changsha 410083, China}

\author{Hao Liu}
\affiliation{School of Physics, Central South University, Changsha 410083, China}

\author{Qi-Yi Wu}
\affiliation{School of Physics, Central South University, Changsha 410083, China}

\author{Chen Zhang}
\affiliation{School of Physics, Central South University, Changsha 410083, China}

\author{Xiao-Fang Tang}
\affiliation{Department of Physics and Electronic Science, Hunan University of Science and Technology, Xiangtan 411201,Hunan, China}

\author{Bo Chen}
\affiliation{School of Physics, Central South University, Changsha 410083, China}

\author{Chuan-Cun Shu}
\affiliation{School of Physics, Central South University, Changsha 410083, China}

\author{Hai-Yun Liu}
\affiliation{Beijing Academy of Quantum Information Sciences, Beijing 100085, China}

\author{Yu-Xia Duan}
\affiliation{School of Physics, Central South University, Changsha 410083, China}

\author{Peter M. Oppeneer}
\affiliation{Department of Physics and Astronomy, Uppsala University, Box 516, S-75120 Uppsala, Sweden}

\author{Jian-Qiao Meng}
\email{Corresponding author: jqmeng@csu.edu.cn}\affiliation{School of Physics, Central South University, Changsha 410083, China}

\date{\today}

\begin{abstract}

We investigate the nonequilibrium electronic and lattice dynamics of the charge-density-wave (CDW) compound 1$T$-TiSe$_2$ using ultrafast optical spectroscopy over a wide range of temperatures and pump fluences. We reveal a close relationship between the observed ultrafast dynamical processes and two characteristic temperatures: $T_{\rm CDW}$ ($\sim$202 K) and $T^*$ ($\sim$165 K). Two coherent phonon modes are identified: a high-frequency $A_{1g}$ mode ($\omega_{1}$) and a lower-frequency $A_{1g}$-CDW amplitude mode ($\omega_{2}$). While both modes soften with increasing temperature, in contrast to thermal behavior we observe a pronounced fluence-induced hardening of the CDW amplitude mode on sub-picosecond timescales. This anomalous frequency upshift provides direct evidence for a nonthermal reconstruction of the lattice potential, driven by transient screening of electron-phonon renormalization by the photoexcited carrier plasma. Concomitantly, the excited-state buildup time exhibits an abrupt increase above a well-defined critical fluence below the CDW transition temperature, signaling a qualitative change in carrier relaxation dynamics. The coincidence between phonon hardening and the fluence threshold indicates that ultrafast electronic screening reshapes the effective lattice potential underlying the CDW order, promoting a nonequilibrium metallic-like response without thermal melting. Our results establish ultrafast phonon hardening as a sensitive probe of lattice potential reconstruction and highlight the fragile balance between excitonic correlations and lattice dynamics in photoexcited 1$T$-TiSe$_2$.
\\
\\
\end{abstract}

\pacs{71.45.Lr, 78.47.J-, 64.70.Rh}

\maketitle

\section{1 Introduction}	

Layered transition metal dichalcogenides (TMDCs), with the general formula M$X_2$ (M = transition metal, $X$ = chalcogenide), are a prominent class of two-dimensional (2D) materials. Their diverse properties, including superconductivity \cite{HNoh2017, SManzeli2017, AJindal2023} and charge density waves (CDW) \cite{KRossnagel2011, JHwang2024}, arise from complex interactions among charge, lattice, spin, and orbital degrees of freedom. This makes TMDCs an excellent platform for investigating correlated electronic systems, particularly under nonequilibrium conditions where the delicate balance among competing interactions can be transiently altered. Among them, 1$T$-TiSe$_2$ is particularly notable as a semimetal with a small band overlap that undergoes a $2\times2\times2$ commensurate CDW transition near 202 K ($T_{\rm CDW}$), accompanied by significant lattice and electronic structural reshaping.

1$T$-TiSe$_2$ exhibits a rich variety of physical phenomena, including CDW formation \cite{Salvo1976}, chiral CDW states \cite{DWickramaratne2022}, and potential excitonic Bose-Einstein condensation \cite{FXBronold2006}, along with superconductivity under doping or pressure \cite{EMorosan2006, AFKusmartseva2009, YIJoe2014, LJLi2016}. At room temperature, 1$T$-TiSe$_2$ is a semimetal, featuring a Se 4$p$ hole pocket at the $\Gamma$ point and Ti 3$d$ electron pockets near the Brillouin zone's M/K points, with an indirect band overlap of approximately 80 meV [Fig.\ \blue{1(a)}] \cite{KRossnagel2002, CMonney2010, MDWatson2019, MHuber2024, PChen2016SR}. Cooling below $T_{\rm CDW}$ induces a commensurate CDW phase with a $2\times2\times2$ periodic lattice distortion \cite{Salvo1976}. This transition has been widely discussed in terms of excitonic insulator behavior, characterized by band folding and partial gap opening at the M/$\Gamma^*$ point, as shown in Fig.\ \blue{1(b)} \cite{SKoley2014, Cercellier2007}.

The precise origin of the CDW phase in 1$T$-TiSe$_2$ remains a subject of ongoing debate, with leading theories considering a combination of Peierls instability, Jahn-Teller effects, and, significantly, the condensation of excitons --- bound electron-hole ($e$-$h$) pairs \cite{EMorosan2006, Cercellier2007, MPorer2014, JWezel2010, AKogar2017, YCheng2022, Hedayat2019}. This last mechanism suggests 1$T$-TiSe$_2$ may realize an excitonic insulator. Further complexity emerges below $T_{\rm CDW}$, where a broad resistivity anomaly appears around 165 K ($T^*$) \cite{Salvo1976, KRossnagel2002}. Despite extensive studies, the microscopic origin of this temperature scale remains unsettled, with proposed interpretations ranging from effective mass renormalization  \cite{CMonney2010} and Fermi surface reconstruction \cite{PKnowles2020} to incoherence-to-coherence crossover \cite{OYi2024} and changes in dominant carrier type \cite{HUeda2021}. These unresolved issues suggest that competing electronic and lattice interactions evolve on multiple energy and time scales, motivating time-resolved probes capable of disentangling their dynamical interplay \cite{Giannetti2016}.

\begin{figure}[b]
\vspace*{-0.2cm}
\begin{center}
\includegraphics[width=0.98\columnwidth,angle=0]{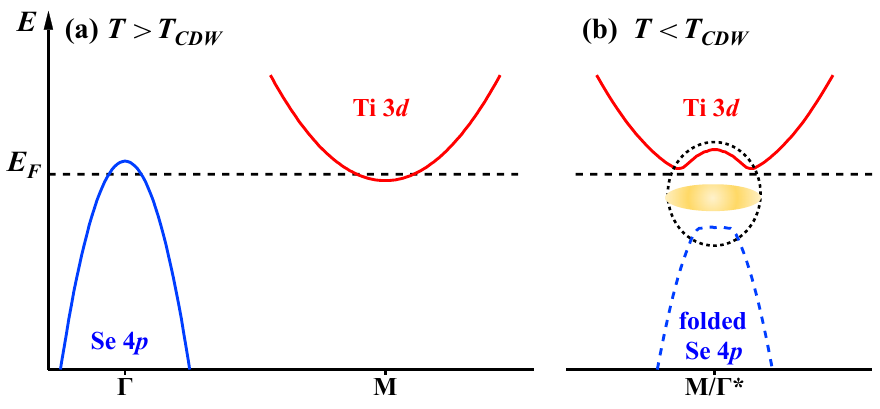}
\end{center}
\vspace*{-0.5cm}
\caption{\label{FIG:1}(a) Electronic structure of 1$T$-TiSe$_2$ along $\Gamma$-M in the normal state ($T > T_{\rm CDW}$). (b) Band structure near M/$\Gamma^*$ in the CDW state ($T < T_{\rm CDW}$), illustrating band folding and the subsequent hybridization between the folded Se 4$p$ and Ti 3$d$ bands, leading to the formation of excitonic correlations between electron-hole pairs.}
\end{figure}

\begin{figure*}[tbp]
\vspace*{-0.2cm}
\begin{center}
\includegraphics[width=1.9\columnwidth]{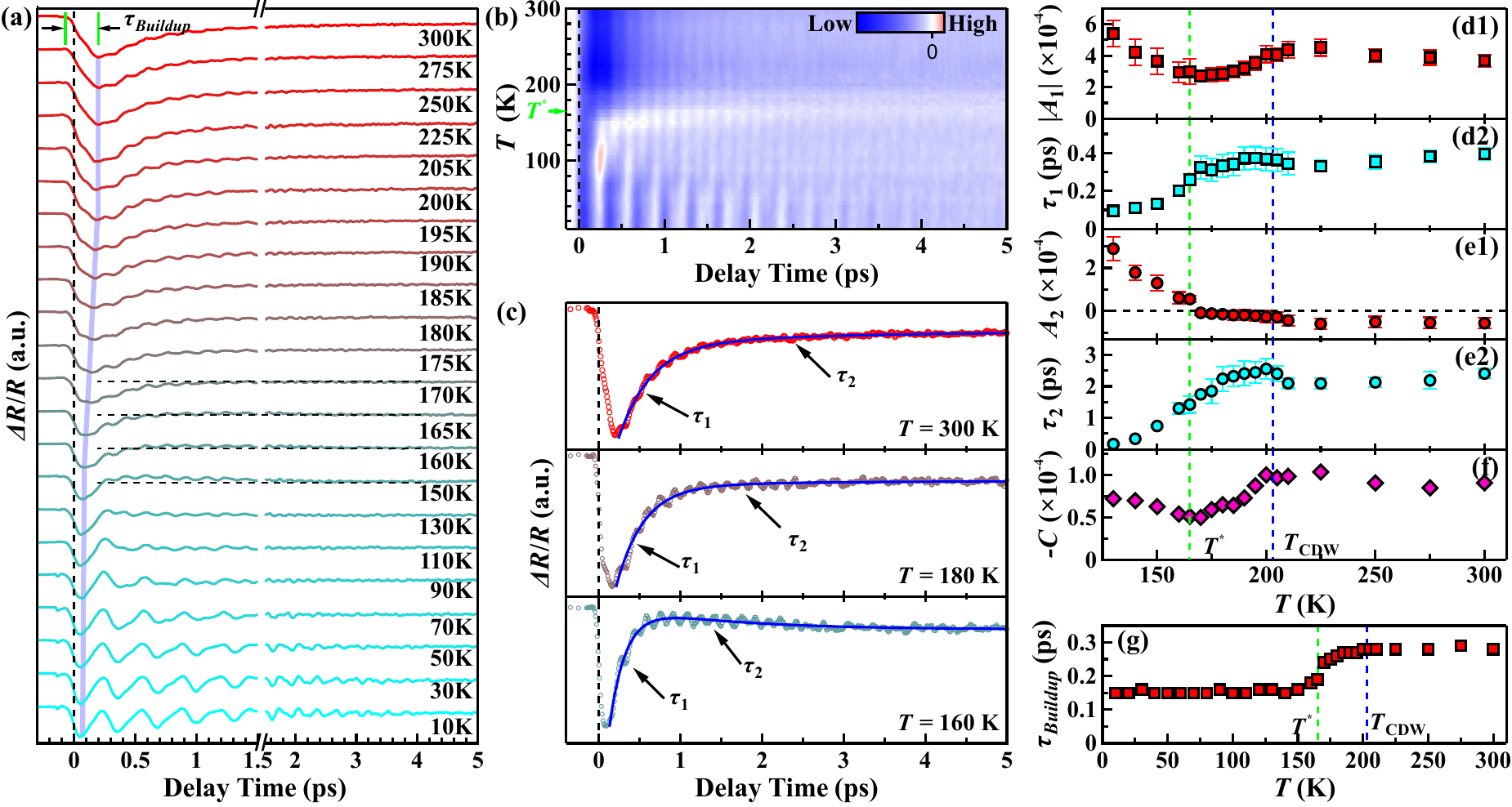}
\end{center}
\vspace*{-0.5cm}
\caption{\textbf{Temperature-dependent ultrafast reflectivity dynamics of 1$T$-TiSe$_2$}. (a) Transient $\Delta R/R$ vs. delay time over temperatures ranging from 10 to 300 K at a pump fluence of $\sim$40 {\mjcm}. Note the break in the $x$ axis. (b) 2D pseudocolor map of $\Delta R/R$ as a function of delay time and temperature. (c) Bi-exponential fits (blue lines) to $\Delta R/R$ at selected temperatures. Arrows indicate the initial ($\tau_1$) and second ($\tau_2$) relaxation processes. (d1, d2) Amplitude ($|A_1|$) and lifetime ($\tau_1$) of the initial relaxation as a function of temperature. (e1, e2) Amplitude ($A_2$) and lifetime ($\tau_2$) of the second relaxation as a function of temperature. (f) Temperature dependence of $C$. (g) Temperature dependence of buildup time ($\tau_{Buildup}$) as defined in (a). The vertical blue and green dashed lines indicate $T_{\rm CDW}$ and $T^*$, respectively.}
\label{FIG:2}
\end{figure*}\setlength{\parskip}{0pt}

In this work, we present a comprehensive ultrafast optical spectroscopy investigation of 1$T$-TiSe$_2$ that resolves the interplay between electronic and lattice dynamics across both temperature and excitation density. By systematically tracking quasiparticle relaxation dynamics, we establish clear correlations with the characteristic temperatures $T_{\rm CDW}$ and $T^*$ ($\sim$165 K), including a pronounced change in carrier relaxation pathways below $T^*$. In addition, we identify two coherent phonon modes: a high-frequency $A_{1g}$ mode ($\omega_1$) and a low-frequency $A_{1g}$-CDW amplitude mode ($\omega_2$). Remarkably, the CDW amplitude mode exhibits an anomalous hardening with increasing pump fluence. This behavior stands in stark contrast to thermal suppression of the CDW order, where phonon softening is expected, and indicates a nonthermal modification of the effective CDW potential. We attribute this hardening to the screening of long-range electron-phonon interactions by the dense photoexcited carrier population, which partially suppresses the electronic renormalization responsible for the lattice instability and drives the phonon frequency toward that of the unrenormalized lattice. Furthermore, we observe an abrupt increase in the excited-state buildup time above a critical fluence, accompanied by characteristic changes in the short-time Fourier spectra. These features signal the emergence of a strong-excitation non-equilibrium regime with suppressed CDW recovery, consistent with recent time-resolved ARPES and diffraction studies that report photoinduced CDW reorganization and domain-wall-dominated dynamics in 1$T$-TiSe$_2$. Our results thus provide a macroscopic optical perspective on the ultrafast, fluence-tunable competition between electronic correlations and lattice dynamics in this prototypical CDW material.

\section{2 Experimental details}

Ultrafast optical spectroscopy measurements were conducted using a noncollinear pump-probe setup driven by a Yb-based femtosecond laser (1 MHz). The output seeded an optical parametric amplifier and compressor, generating $\sim$35 fs pulses centered at 800 nm (1.55 eV) with a bandwidth of $\sim$35 nm \cite{CZhang2022, BLTan2025}. Orthogonally polarized pump ($\sim$80 $\mu$m spot) and probe ($\sim$40 $\mu$m spot)  beams were focused onto freshly cleaved 1$T$-TiSe$_2$ single crystals, grown by chemical vapor transport with iodine \cite{XFTang2022}. Experiments were performed under high vacuum (10$^{-6}$ Torr) across a temperature range of $4-300$ K.

\section{3 Results and discussion}

Figure \blue{2(a)} presents the transient reflectivity ($\Delta R/R$) of 1$T$-TiSe$_2$ from 10 to 300 K, measured at a low pump fluence (40 {\mjcm}), well below the critical fluence for quenching exciton condensation \cite{SFDuan2021}. Photoexcitation elicits an instantaneous $\Delta R/R$ change followed by multiple recovery processes, indicating multiple relaxation channels governed by the CDW-modified electronic structure. A strong temperature dependence is evident, marked by pronounced oscillations throughout the entire range. The time to reach the initial $\Delta R/R$ minimum shifts to longer delays with increasing temperature (purple line), a behavior discussed further below.

As temperature decreases, the oscillations' amplitude and contribution significantly increase, dominating the low-temperature signal. The 2D pseudocolor map in Fig.\ \blue{2(b)} further illustrates this, revealing a robust, long-period oscillation at low temperatures. As shown in the time-domain data of Fig.\ \blue{2(a)}, this oscillation exhibits a cosine-like form, a signature of a displacive excitation. This mechanism involves the sudden modification of the ionic potential-energy surface following photoexcitation \cite{CMonney2016}.This long-period mode gradually damps and diminishes with increasing temperature, merging with a shorter-period oscillation around 130 K. Above 130 K, only weaker, shorter-period oscillations are resolvable up to 300 K. Notably, at high temperatures, the $\Delta R/R$ signal is entirely negative; however, below $T^*$ (165 K), a positive signal emerges around 0.5 ps delay. This distinct temperature evolution of $\Delta R/R$ highlights its sensitivity to subtle electronic structure changes associated with $T^*$. The strong temperature dependence and displacive character indicate that this mode directly tracks changes in the CDW-related lattice potential.

To quantitatively analyze the observed dynamics, we isolated the non-oscillatory component of the transient reflectivity. As shown by the solid blue lines in Fig.\ \blue{2(c)}, this response is accurately described by a bi-exponential decay model: $\Delta R/R$ = $A_1e^{-t/\tau_1}$ + $A_2e^{-t/\tau_2}$ + $C$. Here, $A_i$ and $\tau_i$ represent the amplitude and relaxation time of the $i$-th component, while $C$ accounts for long-lifetime processes. While phenomenological, this bi-exponential description captures the dominant relaxation timescales consistently across the studied temperature range. This fitting was applied to data after the initial $\Delta R/R$  minimum. The extracted temperature dependence of amplitudes ($|A_1|$ and $A_2$), relaxation times ($\tau_1$ and $\tau_2$) and $C$ are presented in Figs.\ \blue{2(d)} - \blue{2(f)}. Due to the dominance of oscillatory signals at low temperatures, our analysis focuses on $T \geq$ 130 K, a range sufficient as no additional transitions are known below this temperature.

The temperature evolution of both amplitudes and relaxation times reveals two distinct anomalies, centered around $T_{\rm CDW}$ and $T^*$. At high temperatures ($T_{\rm CDW}$), $\tau_1$ slightly decreases with decreasing temperature, consistent with observations in other semimetallic materials like 1$T$-TiTe$_2$ \cite{SXZhu2021}, and WTe$_2$ \cite{YMDai2015}. Given its hundreds-of-femtoseconds timescale, $\tau_1$ can be naturally associated with electron-phonon ($e$-$ph$) thermalization, representing the cooling of the hot electronic system through energy transfer to the lattice. Electron-electron ($e$-$e$) scattering, occurring on much faster timescales, is not the dominant process here. Following this initial thermalization, quasiparticles near the Fermi level undergo further decay. This second component, $\tau_2$, is likely associated with phonon-assisted $e$-$h$ recombination between the conduction (Ti 3$d$) and (folded) valence (Se 4$p$) bands (Fig.\ \blue{1(a)}), a commonly observed process in semimetallic systems \cite{SXZhu2021, YMDai2015}.

As temperature approaches $T_{\rm CDW}$ from above, both $\tau_1$ and $\tau_2$ show a modest increase, signaling the influence of electronic structure modifications due to the impending CDW formation. Upon further cooling below $T_{\rm CDW}$, both amplitudes ($|A_1|$ and $|A_2|$) and lifetimes ($\tau_1$ and $\tau_2$) decrease with decreasing temperature. A more dramatic change occurs when cooling across $T^*$. Below $T^*$, $|A_1|$ increases with decreasing temperature, while $\tau_1$ simultaneously decreases at an accelerated rate. Concurrently, $\tau_2$ continues to decrease, with its rate of reduction slightly increasing as temperature drops. Notably, $A_2$ undergoes a sign change from a small negative to a positive value and rapidly increases in magnitude with further cooling. This profound sign change in $|A_2|$ may reflect a qualitative change in the non-equilibrium relaxation channels, potentially linked to enhanced excitonic correlations that increase the susceptibility to photoinduced domain-wall-like distortions, as discussed in Ref.\cite{SFDuan2021}. As shown in Fig. \blue{2(f)}, $C$ also shows anomalies near both $T_{\rm CDW}$ and $T^*$.

To elucidate the distinct temperature-dependent behavior around $T^*$, we propose two possible mechanisms. First, a change in recombination dynamics may be responsible. Above $T^*$, recombination could primarily involve free $e$-$h$ pairs. However, below $T^*$, the increasing prominence of excitonic correlations or the formation of collective CDW states might significantly alter these recombination pathways. This aligns with previous ARPES studies on 1$T$-TiSe$_2$ reporting an incoherence-to-coherence crossover at 165 K \cite{OYi2024}, attributed to potential exciton condensation below the primary CDW transition. Exciton-assisted recombination can produce an optical response with an opposite sign to that of free-carrier recombination, because excitons interact differently with the probe light: instead of contributing to absorption through free-carrier transitions, bound $e$-$h$ pairs modify the dielectric function via their discrete resonance states and altered selection rules. Second, a shift in the dominant carrier type could contribute. A downshift of the Ti 3$d$ band around 165 K has been observed, leading to a transition from hole-like to electron-like dominant carriers with decreasing temperature \cite{HUeda2021}. Such a change in carrier majority could manifest as a sign change in the $\Delta R/R$ signal, as optical response is sensitive to the density of states and effective mass of the involved carriers. These two scenarios are not mutually exclusive and may jointly contribute to the observed sign reversal across $T^*$.

We next analyze the excited state buildup time, $\tau_{Buildup}$, another notable temperature-dependent quantity. Given the strong oscillatory signals in $\Delta R/R$ at low temperatures (Figs. \blue{2(a)} and \blue{2(b)}), standard fitting methods for $\tau_{Buildup}$ are challenging \cite{BLTan2025, MHuber2022}. We therefore define $\tau_{Buildup}$ as the interval from the onset of reflectivity change until the signal reaches its ``global minimum", as schematically shown in Fig. \blue{2(a)}. This definition does not account for instrumental time resolution, implying the reported $\tau_{Buildup}$ will be slightly longer than the intrinsic excited state formation time. Figure \blue{2(g)} shows the temperature dependence of $\tau_{Buildup}$. Below $T^*$, $\tau_{Buildup}$ remains nearly constant at $\sim$0.15 ps. Upon warming across $T^*$, $\tau_{Buildup}$ increases and reaches a new plateau of $\sim$0.3 ps above $T_{\rm CDW}$. This increase, initiated at $T^*$, reflects a gradual thermal weakening of the stabilizing CDW correlations, which alters the carrier relaxation dynamics even while the CDW order remains intact.As the temperature exceeds $T_{\rm CDW}$, long-range order vanishes, yet short-range fluctuations and soft phonon modes continue to persist. In this high-temperature regime, the suppression of excitonic correlations causes the transient response to shift from being electronically driven to being dominated by lattice dynamics. Consequently, because structural processes typically evolve on slower timescales than purely electronic ones, the system exhibits a relatively longer buildup time. Indeed, the $\sim$0.3 ps buildup time measured in the normal state is markedly longer than those of conventional semimetals \cite{SXZhu2021, YMDai2015}, reinforcing the view that initial excited-state formation in 1$T$-TiSe$_2$ is governed by complex collective interactions rather than simple, independent electronic transitions.

At low temperatures ($T < T^*$), the $\sim$0.15 ps excited state buildup time is comparable to the timescale for CDW gap reduction (quenching) observed in Tr-ARPES measurements at low pump fluence \cite{MHuber2022, SMathias2016, SFDuan2023}. This correlation is consistent with a scenario in which carrier multiplication via interband impact ionization, involving scattering between the Ti 3$d$ and Se 4$p$ bands, plays an important role in governing the excited-state buildup dynamics in the low-temperature CDW phase \cite{SMathias2016}. As temperature increases above $T^*$, the CDW gap continues to decrease and the Ti 3$d$ band shifts upward in energy. This expands the phase space for $e$-$e$ interband scattering, prolonging carrier accumulation near the Fermi level and consequently extending $\tau_{Buildup}$. When temperature exceeds $T_{\rm CDW}$, the CDW gap closes, and the characteristic band folding vanishes. In this normal state, optical excitation directly promotes electrons from the Se 4$p$-derived valence band to the conduction band around the $\Gamma$ point. These hot electrons rapidly scatter away from $\Gamma$ and accumulate at the M point within the conduction band. The timescale for these interband scattering and accumulation processes aligns well with the $\sim$0.3 ps $\tau_{Buildup}$ observed in the normal state by Tr-ARPES \cite{CMonney2016}. This agreement highlights that ultrafast optical reflectivity captures the same underlying electronic relaxation processes identified by momentum-resolved probes, while providing a complementary, macroscopic perspective.

Having analyzed the non-oscillatory response, we now turn to the significant oscillatory component of the $\Delta R/R$ signal. These oscillations are indicative of collective bosonic excitations and serve as a reliable measure of phase transitions \cite{HLiu2024, YZZhao2023}. Prominent high-frequency oscillations appear instantaneously upon photoexcitation, superimposed on the $\Delta R/R$ profile (Figs.\ \blue{2(a)} and \blue{2(b)}). The oscillatory components, obtained by subtracting the non-oscillatory decay, are presented in Fig.\ \blue{3(a)} for several selected temperatures. With increasing temperature, the oscillation amplitudes gradually decrease and a clear beating pattern develops, reflecting the coexistence of multiple collective excitation modes with distinct frequencies.

\begin{figure}[tbp]
\vspace*{-0.2cm}
\begin{center}
\includegraphics[width=1\columnwidth]{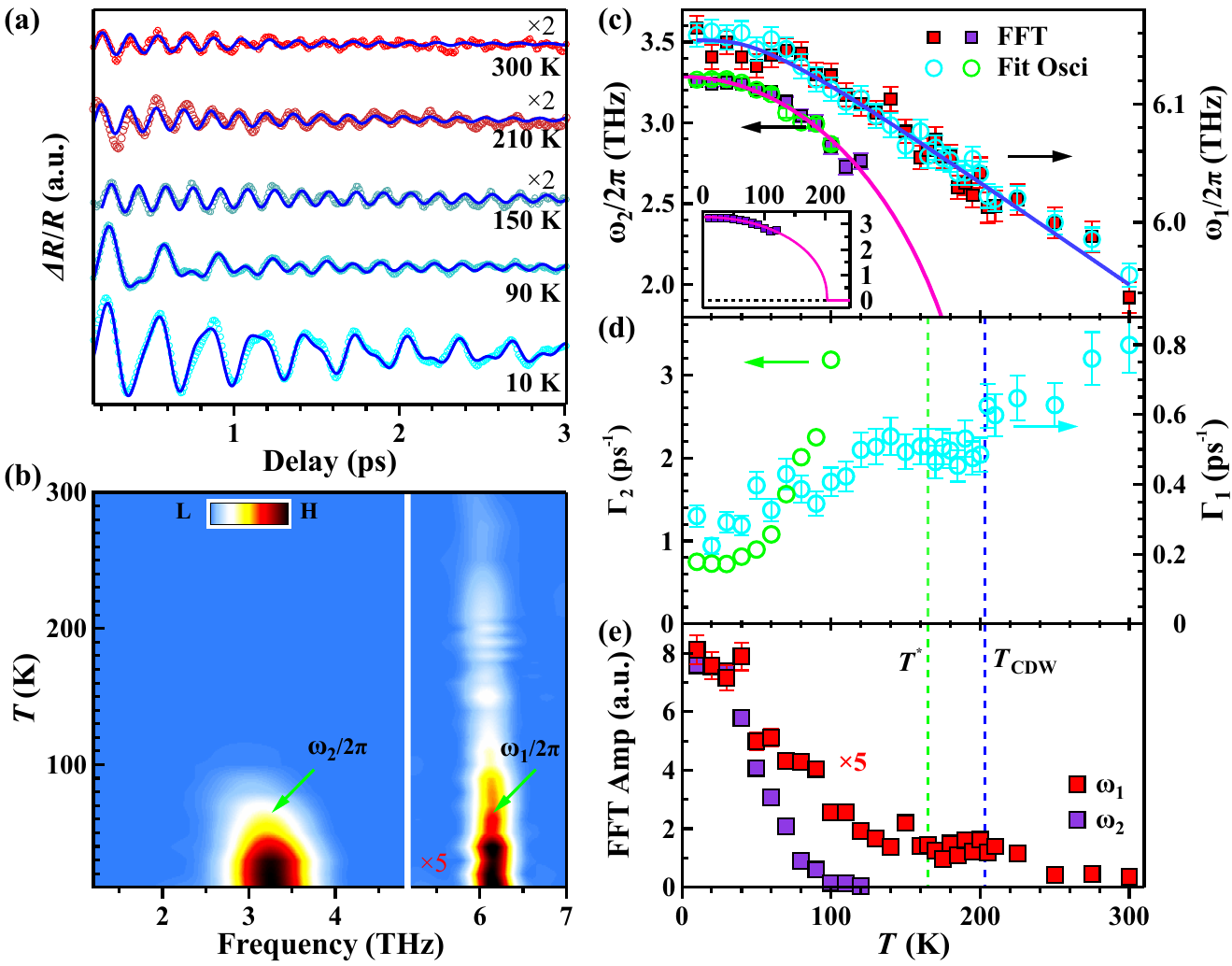}
\end{center}
\vspace*{-0.5cm}
\caption{\label{FIG:3} \textbf{Oscillatory part of temperature-dependent data} (a) Oscillations from the transient reflectivity at five selected temperatures, with blue lines represent fits using Eq.\ (\blue{1}). (b) False-color FFT spectrum as a function of frequency and temperature. (c) Temperature dependence of $\omega_1/2\pi$ and $\omega_2/2\pi$ from the FFT (squares) and fits (circles). The solid magenta lines is a fit to the frequency $\omega_2$ with a mean-field-like as described in the text. The solid cyan line in (c) is anharmonic phonon model fit to data obtained from fitting to Eq.\ (\blue{1}).  (d) Temperature dependence of damping rates $\Gamma_1$ and $\Gamma_2$ from fits. (e) Temperature dependence of FFT amplitudes of $\omega_1$ and $\omega_2$.}
\end{figure}

To further characterize these bosonic excitations, we performed Fast Fourier Transform (FFT) analysis on the oscillatory data. Figure \blue{3(b)} presents the FFT spectrum as a function of frequency and temperature. At low temperatures, two distinct terahertz modes, $\omega_1$ and $\omega_2$, were observed. At 10 K, their frequencies were 6.16 THz (25.5 meV or 205.5 cm$^{-1}$) and 3.26 THz (13.5 meV or 108.7 cm$^{-1}$), respectively. With increasing temperature, the spectral weight of both modes is rapidly suppressed, and $\omega_2$ becomes difficult to resolve above approximately 120 K. The higher frequency mode, $\omega_1$, corresponds to the coherent $A_{1g}$ phonons and persists up to room temperature \cite{XFTang2022, JAHoly1977}. Conversely, the lower frequency mode, $\omega_2$, is attributed to a photoinduced coherent $A_{1g}$-CDW amplitude phonon \cite{MPorer2014, Hedayat2019, SFDuan2021, CSSnow2003}. This mode is a direct consequence of the periodic lattice distortion (PLD) and is absent in the normal state. The extracted frequencies and FFT amplitudes for both modes are summarized in Figs. \blue{3(c)} and \blue{3(e)} (squares).

\begin{figure*}[tbp]
\vspace*{-0.2cm}
\begin{center}
\includegraphics[width=1.8\columnwidth]{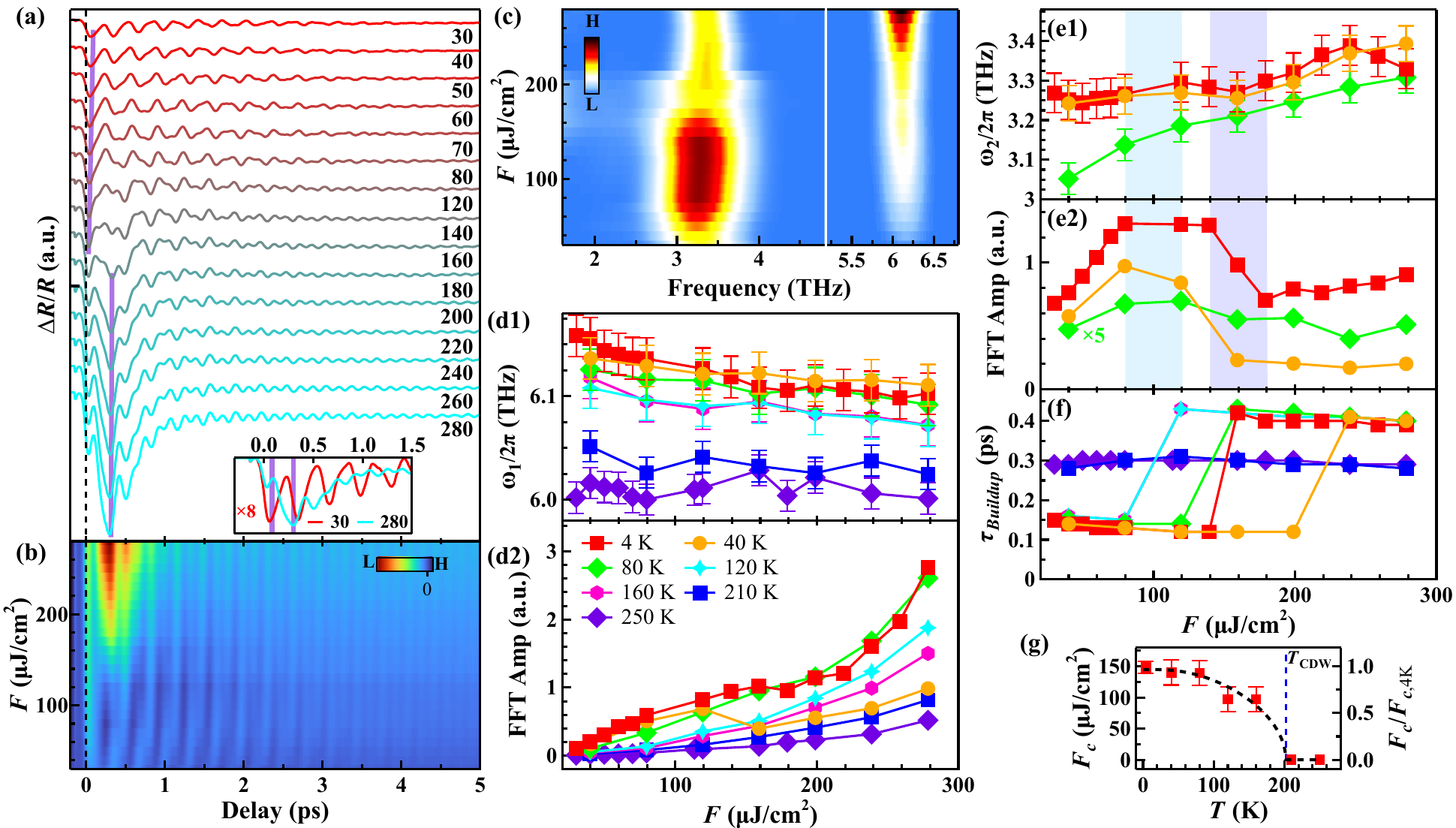}
\end{center}
\vspace*{-0.5cm}	
\caption{\label{FIG:4} \textbf{Pump fluence-dependent ultrafast dynamics of 1$T$-TiSe$_2$.} (a) Transient $\Delta R/R$ as a function of delay time for various pump fluences at 4 K. (b) 2D pseudocolor map of $\Delta R/R$ as a function of delay time and pump fluence. (c) False-color FFT spectrum as a function of frequency and pump fluence. (d1,d2) Fluence dependence of $\omega_1$ frequency and FFT amplitude at different temperatures. (e1, e2) Fluence dependence of $\omega_2$ frequency and FFT amplitude at different temperatures. (f) Fluence dependence of $\tau_{Buildup}$ for various temperatures. (g) Temperature dependence of critical fluence $F_c$. The dashed black line is a fit to the $F_c(T)$ using an empirical mean-field expression for a second-order phase transition, $F_c(T) \propto$ tanh$^2[\alpha \sqrt{T_{\rm CDW}/T - 1}\Theta(T_{\rm CDW}-T)]$ \cite{ZTLiu2023}, where $\alpha$ is a fitting parameter and $\Theta$ is the Heaviside step function.}
\end{figure*}

Phonon properties were also extracted by fitting the damped oscillations using the following expression [blue solid lines in Fig.\ \blue{3(a)}]:
\begin{equation}
(\frac{\Delta R}{R})_{osc}= \sum\limits_{i=1,2} A_{i}e^{-\Gamma_i t}{\rm sin}(\omega_{i} t + \phi_{i}),
\label{eqOSCI}
\end{equation}
where $A_i$, $\Gamma_i$, $\omega_i$, and $\phi_i$ are the amplitude, damping rate, frequency, and initial phase of the $i$-th oscillatory signal, respectively. The extracted temperature dependence of $\omega_1$ and $\omega_2$ is plotted in Fig.\ \blue{3(c)} (circles), and damping rates $\Gamma_1$ and $\Gamma_2$ in Fig.\ \blue{3(d)} (circles). All parameters exhibit significant variations with temperature. Specifically, both $\omega_1$ and $\omega_2$ soften (redshift) as temperature increases. The temperature evolution of $\omega_1$ is well described by conventional anharmonic phonon decay within the experimental resolution \cite{MBalkanski1983, JMenendez1984}. In contrast, the damping rate $\Gamma_1$ shows pronounced anomalies near $T_{\rm CDW}$ and $T^*$, and the FFT amplitude of $\omega_1$ also displays distinct anomalies at these characteristic temperatures (Fig. \blue{3(e)}), indicating an unexpected sensitivity of the $A_{1g}$ phonon to the electronic and lattice reconstruction associated with the CDW transition and the $T^*$ anomaly.

The lower-frequency mode $\omega_2$ exhibits hallmark characteristics of a CDW amplitude mode. Its temperature dependence follows a mean-field-like behavior \cite{WKLee1988, LCui2017}: $\omega_2$ = $\omega_2(0) \times(1-T/T_{\rm CDW})^\beta$, with $\beta$ = 0.22, smaller than the mean-field value of 0.5 but consistent with the experimentally determined $T_{\rm CDW}$ = 202 K. In addition, the spectral weight of $\omega_2$ diminishes as temperature approaches $T_{\rm CDW}$, reflecting the gradual loss of long-range CDW coherence. Together, these observations establish $\omega_2$ as a direct probe of the stability and renormalization of the CDW lattice potential.

The interplay between low carrier density, weak Coulomb screening, and excitonic correlations in 1$T$-TiSe$_2$ makes this system particularly susceptible to photoinduced perturbations. Introducing excess carriers through ultrafast photoexcitation enhances screening and scattering processes, thereby modifying the electronic renormalization that stabilizes the CDW state \cite{LStojchevska2014}. Motivated by this sensitivity, we systematically investigated the fluence dependence of the ultrafast dynamics to explore photoinduced non-equilibrium modifications of the CDW state.

Figure \blue{4(a)} presents the transient differential reflectivity ($\Delta R/R$) as a function of pump-probe delay for various pump fluences (from $\sim$30 {\mjcm} to $\sim$280 {\mjcm}) at 4 K. The $\Delta R/R$ response exhibits notable variations, particularly an abrupt increase in the excited state buildup time ($\tau_{Buildup}$) above a critical fluence $F_c$ $\approx$ 140 {\mjcm}. This fluence scale is comparable to that reported previously for photoinduced suppression of CDW order in 1$T$-TiSe$_2$ \cite{SFDuan2023}. The complex interplay of photoinduced effects is further visualized in the 2D pseudocolor map of $\Delta R/R$ in Fig.\ \blue{4(b)}, which spans pump-probe delay and fluence. This map clearly reveals multiple distinct oscillatory features, all exhibiting strong fluence dependence. FFT analysis of these data (Fig. \blue{4(c)}) confirms the presence of the two phonon modes $\omega_1$ and $\omega_2$ over the entire fluence range.

Figures \blue{4(d1)} and \blue{4(d2)} display the fluence dependence of the frequency and FFT amplitude of $\omega_1$ at various temperatures. As expected \cite{SXZhu2021}, the frequency of $\omega_1$ generally decreases with increasing pump fluence, indicative of photoinduced softening. This softening is substantially more pronounced below $T_{\rm CDW}$ than in the normal state, underscoring the strong coupling between $\omega_1$ and the CDW order. The FFT amplitude of $\omega_1$ increases with fluence but exhibits non-monotonic behavior at low temperatures, suggesting additional modifications of the lattice response under strong excitation.

In stark contrast, the CDW amplitude mode $\omega_2$ displays an anomalous fluence dependence. Figures \blue{4(e1)} and \blue{4(e2)} show the frequency and amplitude of $\omega_2$ as a function of fluence at 4 K, 40 K, and 80 K. Surprisingly, unlike conventional materials \cite{CZhang2022, SXZhu2021, QYWu2025}, $\omega_2$ exhibits a pronounced frequency hardening (blueshift) with increasing pump fluence. This effect cannot be attributed to laser heating, which is known to downshift the mode frequency [Fig.\ \blue{3(c)}]. Systematic short-time Fourier transform (STFT) analysis reveals that this upshift initiates at low excitation densities and evolves continuously, indicating that the nonthermal reconstruction of the lattice potential precedes the complete collapse of the CDW order. The magnitude of this hardening is particularly striking at intermediate temperatures. Specifically, at 80 K, $\omega_2$ monotonically increases by approximately 8\% up to $\sim$280 {\mjcm}, a rate twice as high as that observed at 4 K. This enhanced hardening reflects the increased susceptibility of the CDW state at 80 K; while the order remains well-established, thermal fluctuations partially weaken the stabilizing electronic correlations, enabling photoexcited carriers to more effectively screen the lattice renormalization. In contrast to the monotonic behavior at 80 K, the response at 4 K is more complex and non-monotonic. At this base temperature, the $\omega_2$ frequency initially increases slowly, dips around 120 {\mjcm}, and then resumes its increase above 160 {\mjcm} before final suppression. Concurrently, the $\omega_2$ amplitude at 4 K exhibits a similar anomaly, decreasing significantly above 120 {\mjcm} before recovering at higher fluences. Similar variations are also observed at 40 K around 160 {\mjcm}. This non-monotonic evolution in both frequency and amplitude likely reflects a CDW phase-reversal or inversion scenario \cite{SFDuan2025}, where the collective response reaches a minimum as the equilibrium order vanishes and the system transitions into a light-induced metastable state.

Short-time Fourier transform (STFT) analysis provides further insight into this behavior. As shown in Fig. \ \blue{S2} of the Supplemental Material \cite{SUPPM}, the frequency upshift of $\omega_2$ occurs essentially instantaneously after photoexcitation and relaxes back toward its equilibrium value within approximately 1 ps at high fluence. This ultrafast response demonstrates that the hardening of $\omega_2$ is a genuinely non-thermal effect and reflects a transient stiffening of the CDW lattice potential. The hardening of the $\omega_{2}$ amplitude mode provides a unique window into the microscopic forces stabilizing the CDW. In the equilibrium CDW state, $\omega_{2}$ acts as the ``soft mode", its frequency significantly suppressed (renormalized) relative to the bare lattice frequency by strong $e$-$p$h coupling and excitonic correlations. Thermal excitations typically flatten the free-energy potential, further softening the mode. However, our observation of fluence-dependent hardening points to a fundamentally nonthermal mechanism, consistent with an ultrafast recovery of the lattice potential toward its bare form.

We attribute this to the transient suppression of the electronic renormalization. The dense photoexcited $e$-$h$ plasma efficiently screens Coulomb interactions and weakens the susceptibility enhancement responsible for the Peierls-like instability. With this electronic ``drag" removed, the effective lattice potential steepens, and the phonon frequency blueshifts toward the stiffer value of the unrenormalized state. This process is not merely a ``decoupling" \cite{KIshioka2008} but a transient removal of the Peierls driving force. The fact that this hardening occurs on a sub-picosecond timescale (as shown by STFT analysis in section \blue{II} in Supplementary Materials \cite{SUPPM}) confirms that the electronic glue holding the CDW distortion is quenched faster than the lattice can thermally relax, effectively snapping the lattice potential back to its metallic stiffness. This interpretation is consistent with time-resolved ARPES and ultrafast electron diffraction studies \cite{SFDuan2021, SFDuan2023}, which have reported photoinduced suppression of CDW order and the emergence of a long-lived non-equilibrium metallic-like state in 1$T$-TiSe$_2$ under comparable excitation conditions. Within this framework, the present optical results provide complementary information by directly tracking the transient reconstruction of the lattice potential through the CDW amplitude mode.

Supporting this picture of potential reconstruction, the excited state buildup time ($\tau_{Buildup}$) exhibits a sharp transition that mirrors the phonon dynamics. As shown in Fig. \blue{4(f)}, below $T_{\rm CDW}$, $\tau_{Buildup}$ undergoes a step-like increase from $\sim$0.15 ps to $\sim$0.4 ps once the fluence exceeds a temperature-dependent critical value $F_c$ (the detailed determination of the critical fluence at 40 K is provided in Section \blue{III} of the Supplemental Material \cite{SUPPM}). This abrupt jump contrasts with the gradual temperature-induced crossover observed in thermal equilibrium, which saturates at a shorter timescale of $\sim$0.3 ps. Instead, the step-like behavior above $F_c$ signifies a genuine nonequilibrium threshold process, where intense photoexcitation rapidly suppresses the CDW gap and drives the system into a metastable metallic state. In this strong-excitation regime, an early local minimum appears in Fig.  \blue{4(a)} prior to the global minimum. This indicates that while the initial electronic response, which is associated with the prompt destruction of excitonic correlations, remains nearly instantaneous, the subsequent population buildup near $E_F$ becomes substantially delayed once the collective CDW order is suppressed. This transition is governed by a temperature-dependent critical fluence $F_c$, which decreases monotonically and vanishes near $T_{\rm CDW}$ [Fig. \blue{4(g)}]. The existence of $F_c$ signifies a threshold beyond which the CDW free-energy landscape is fundamentally modified, allowing the system to access a nonequilibrium regime with altered carrier relaxation dynamics. Specifically, the collapse of the CDW gap above $F_c$ restores a metallic-like phase space, triggering a shift from the rapid carrier multiplication via impact ionization dominant in the gapped phase to multiple $e$-$e$ scattering channels. This redistribution of carrier energy slows the accumulation process near $E_F$, thereby prolonging the population buildup to $\sim0.4$ ps.

In the low-fluence CDW phase, the presence of the gap restricts the available phase space, and carrier multiplication via impact ionization dominates, giving rise to short buildup times. Above $F_c$, the strong suppression of CDW order closes the gap and restores a metallic-like electronic structure, leading to prolonged carrier accumulation and relaxation. The coincidence between the threshold behavior in $\tau_{Buildup}$ and the hardening of $\omega_2$ demonstrates that this non-equilibrium regime is characterized by a transiently stiffened interatomic potential liberated from CDW-induced softening. Thus, the photoinduced response of 1$T$-TiSe$_2$ reflects not a simple thermal melting of the CDW, but a deterministic, ultrafast reconstruction of the coupled electronic-lattice system driven by carrier-induced screening.

\section{4 Conclusions}

In conclusion, ultrafast optical spectroscopy shows that strong photoexcitation drives 1$T$-TiSe$_2$ far from equilibrium and induces a sharp, fluence-dependent reorganization of its coupled electronic and lattice dynamics. This nonthermal response is characterized by a threshold fluence, an abrupt increase in the excited-state buildup time, and a sub-picosecond hardening of the $A_{1g}$-CDW amplitude mode ($\omega_2$). Since $\omega_2$ is tied to the CDW-distorted lattice, its transient frequency upshift directly reflects a rapid stiffening of the lattice potential caused by photoinduced screening of electronic renormalization. The coincidence of these signatures indicates a strong suppression of CDW-stabilizing interactions, allowing access to a long-lived nonequilibrium regime with weakened CDW correlations. Within this framework, the characteristic temperature $T^*$ naturally corresponds to a crossover involving enhanced excitonic correlations and modified recombination pathways, rendering the low-temperature CDW state particularly sensitive to photoexcitation. Our results demonstrate that excitonic correlations and electron-phonon coupling can be dynamically tuned on ultrafast timescales, leading to a nonthermal reconstruction of the CDW free-energy landscape rather than simple thermal melting.

\textit{This work was supported by the National Key Research and Development Program of China (Grant No.\  2022YFA1604204), the Beijing National Laboratory for Condensed Matter Physics (No.\ 2024BNLCMPKF001), the National Natural Science Foundation of China (Grant No.\ 12574168), and the Science and Technology Innovation Program of Hunan Province (Grant No.\ 2022RC3068). P. M. O. acknowledges funding from the K.\ and A.\ Wallenberg Foundation (Grants No.\ 2022.0079 and No.\ 2023.0336) and from the Swedish Research Council (VR) (Grant No.\ 2022-06725).}

\end{document}